\documentclass[12pt]{article}
\usepackage{amsmath,amsthm, amssymb}

\newcommand{\be}{\begin{equation}}
\newcommand{\ee}{\end{equation}}
\newcommand{\ba}{\begin{array}}
\newcommand{\ea}{\end{array}}
\newcommand{\bea}{\begin{eqnarray}}
\newcommand{\eea}{\end{eqnarray}}
\newcommand{\bma}{\begin{matrix}}
\newcommand{\ema}{\end{matrix}}
\newcommand{\bpm}{\begin{pmatrix}}
\newcommand{\epm}{\end{pmatrix}}
\newcommand{\nn}{\nonumber}
\newcommand{\dst}{\displaystyle}

\newcommand{\aone}{\alpha_1}
\newcommand{\qstar}{q_{12}^{\ast}}

\newcommand{\ha}[1]{\frac{#1}{2}}
\newcommand{\sx}{\sqrt6}
\newcommand{\Dh}{\widehat D}
\newcommand{\p}{\partial}
\newcommand{\hf}{{\hat5}}
\newcommand{\effh}{{e_5^\hf}}

\newcommand{\pr}{\prime}
\newcommand{\ov}{\overline}
\newcommand{\wh}{\widehat}
\newcommand{\wt}{\widetilde}
\newcommand{\Psitil}{\wt \Psi}

\newcommand{\Eta}{\mathcal H}
\newcommand{\Etatil}{\wt \Eta}

\newcommand{\G}{\chi_{_G}}
\newcommand{\Gbar}{\ov\chi_{_G}}
\newcommand{\psibar}{\ov \psi}
\newcommand{\phibar}{\ov \phi}
\newcommand{\chibar}{\ov \chi}
\newcommand{\etabar}{\ov \eta}
\newcommand{\sibar}{\ov \sigma}

\newcommand{\ep}{\varepsilon}
\newcommand{\eps}{\epsilon}
\newcommand{\al}{\alpha}
\newcommand{\la}{\lambda}
\newcommand{\da}{\delta}
\newcommand{\om}{\omega}
\newcommand{\Ga}{\Gamma}
\newcommand{\Si}{\Sigma}
\newcommand{\si}{\sigma}

\title{\bf Brane-Localized Goldstone Fermions\\
in Bulk Supergravity}
\author{\\[1cm]
 \large{\bf Jonathan A.~Bagger}~\thanks{bagger@jhu.edu} 
 \quad \it{and} \quad    
 \large{\bf Dmitry V.~Belyaev}~\thanks{belyaev@pha.jhu.edu}\\[1cm]
 \it Department of Physics and Astronomy,\\ 
 \it The Johns Hopkins University,\\
 \it 3400 North Charles Street,\\
 \it Baltimore, MD 21218, USA}
\date{}

\begin{document}

\numberwithin{equation}{section}

\maketitle
\begin{abstract}
We construct the action and transformation laws for bulk five-dimensional AdS supergravity coupled to one or two brane-localized Goldstone fermions.  The resulting bulk-plus-brane system gives a model-independent description of brane-localized supersymmetry breaking in the Randall-Sundrum scenario.  We explicitly reduce the action and transformation laws to spontaneously broken four-dimensional\break supergravity.
\end{abstract}

\newpage


\section{Introduction}

In the Randall-Sundrum scenario \cite{rs}, spacetime is a slice of AdS$_5$, with cosmological constant $\lambda$, bounded by three-branes with tensions $\lambda_1$ and $\lambda_2$ (we follow notation of Ref.~\cite{bb1}).  The setup can be made supersymmetric when the tensions are tuned \cite{first}, with $\lambda_1= \lambda_2= \pm\lambda$, and even when they are not \cite{bb1}, provided $|\lambda_{1,2}|< \lambda$.  In the tuned case, the low-energy effective theory is four-dimensional supergravity with no cosmological constant.  In the detuned case, the effective theory is four-dimensional supergravity with a negative cosmological constant.

In any supersymmetric theory, it is important to investigate ways in which supersymmetry can be broken spontaneously.  In ref.~\cite{bb2} we showed that Scherk-Schwarz mechanism offers one possibility, but only in the detuned case.  The Scherk-Schwarz supersymmetry breaking parameter is the difference between the phases of $\al_1$ and $\al_2$, the coefficients of the brane-localized gravitino mass terms.  In ref.~\cite{br} it was shown that this order parameter is equivalent to a VEV for $B_5$ (the fifth component of the graviphoton in supergravity multiplet). 

In this paper we show that bulk-plus-brane supersymmetry can also be broken by brane-localized fields, whether or not the tensions are tuned.  We start by assuming that supersymmetry is spontaneously broken by brane-localized dynamics.  The precise mechanism is not important; what is essential is that the supersymmetry breaking gives rise to a brane-localized Goldstone fermion.  We require that such a Goldstone fermion exists on one or both branes; we ignore all other brane-localized fields associated with the supersymmetry breaking.  As usual, the Goldstone fermions transform nonlinearly under supersymmetry. 

Let $v_1$ and $v_2$ denote the scales of supersymmetry breaking on the two branes.  In section~\ref{LocalSusy} of this paper, we couple the nonlinearly transforming brane-localized Goldstone fermions to five-dimensional bulk supergravity.  Local supersymmetry imposes a relation between $\la$, $\la_1$, $\al_1$, and $v_1$ (and similarly between $\la$, $\la_2$, $\al_2$, and $v_2$). This relaxes the condition found in ref.~\cite{bb1}.  In particular, with Goldstone fermions on the branes, the bulk-plus-brane action with $|\la_{1,2}|>\la$ can also be made locally supersymmetric. The effective theory for this case is four-dimensional supergravity with a positive cosmological constant.

In section \ref{DimRed}, we reduce the system to four dimensions.  We identify the low-energy degrees of freedom, write the dependence on the fifth coordinate in terms of warp factors, and find a system of equations for the warp factors.  These equations, together with corresponding boundary conditions, determine the supersymmetry breaking in the effective theory.  In section~\ref{Example} we compute the supersymmetry breaking in the tuned Randall-Sundrum scenario. Conventions and supplementary material are collected in a series of appendices.


\section{Local supersymmetry}
\label{LocalSusy}

In this section we construct a supersymmetric bulk-plus-brane system consisting of five-dimensional bulk supergravity, with cosmological constant $\lambda$, compactified on an $S^1/\mathbb{Z}_2$ orbifold.  We place three-branes $\Sigma_i$ of tension $\lambda_i$, with $i=1,2$, at the orbifold fixed points.  We include Goldstone fermions on the branes, remnants of brane-localized supersymmetry breaking dynamics.  We proceed step by step, first considering the bulk, and then adding the branes, one at a time.

\subsection{Bulk action}

We start with the bulk action, as described in ref.~\cite{bb1}.  The action is given by
\bea
\label{bulk}
S_{\rm bulk} = \int d^5\!x e_5 
\Big\{
-\frac{1}{2} R + 6\la^2 
+\frac{i}{2}\Psitil_M^i\Ga^{MNK} D_N\Psi_{Ki}
-\frac{3}{2} \la\, \vec q\cdot \vec\si_i{}^j \Psitil_M^i\Si^{MN}\Psi_{Nj} 
\nn\\
-\frac{1}{4}F_{MN}F^{MN} 
- i\frac{\sqrt6}{16}F_{MN}
\left(
2\Psitil^{Mi} \Psi_i^N + \Psitil_P^i\Ga^{MNPQ}\Psi_{Qi} 
\right) 
\nn\\
-\frac{1}{6\sqrt6}\eps^{MNPQK} F_{MN}F_{PQ}B_K
+\frac{\sqrt6}{4} \la\, \vec q\cdot \vec\si_i{}^j
B_N\Psitil_M^i\Ga^{MNK}\Psi_{Kj}
\Big\}.
\eea
Here $\la$ is a mass parameter, determining the bulk cosmological constant, $\Lambda_5=-6\la^2 k_5^{-2}$ (we set $k_5=1$); $\vec q=(q_1, q_2, q_3)$ is a dimensionless unit vector, characterizing the gauged $U(1)$ of the $SU(2)$ R-symmetry group. The action is invariant, up to boundary terms, under the following supersymmetry transformations,
\bea
\label{bulksusy}
\da e_M^A &=& i \Etatil^i\Ga^A\Psi_{Mi} \\
\da B_M &=& i\frac{\sqrt6}{2}\Psitil_M^i\Eta_i \\
\da \Psi_{Mi} &=& 2 \big(
D_M\Eta_i - i\frac{\sqrt6}{2}\la\, \vec q\cdot \vec\si_i{}^j B_M\,\Eta_j \big)
+i\la\, \vec q\cdot \vec\si_i{}^j \,\Ga_M \,\Eta_j \nn\\
&& +\frac{1}{2\sqrt6} \left(
\Ga_{MNK} - 4g_{MK}\Ga_N \right) F^{NK}\Eta_i.
\eea
In the rest of this work, we use two-component spinors $(\eta_1, \eta_2)$, $(\psi_{m1}, \psi_{m2})$ and $(\psi_{51}, \psi_{52})$, which are constituents of the symplectic Majorana spinors $\Eta_i$, $\Psi_{mi}$ and $\Psi_{5i}$, respectively.  The rules for passing between the two notations, as well as other conventions, are described in ref.~\cite{bb1}.

\subsection{One brane} 

We work in the ``upstairs'' picture, on the covering space of the orbifold. (In this picture, ``boundary'' or ``total derivative'' terms can be neglected.)  For now, we consider just one brane $\Si$, located at $x^5 \equiv z=0$.  As in ref.~\cite{bb1}, we choose the following parity assignments
\bea
\label{paras}
\bma
\text{even}:&\p_m&e_m^a&\effh&B_5&
\eta_1&\psi_{m1}&\psi_{52}&q_{1,2}&{\la} \\
\text{ odd}:&\p_5&e_m^{\hf} & e_5^a&B_m&
\eta_2&\psi_{m2}&\psi_{51}&q_3 ,&{}
\ema
\eea
and redefine the odd fields by explicitly separating out the sign function,
$\ep(z)$,
\bea
\label{redef}
\bma
e_m^\hf  \rightarrow \ep(z) \, e_m^\hf, &
e_5^a \rightarrow \ep(z) \, e_5^a, &
B_m \rightarrow \ep(z) \, B_m \\[2mm]
\eta_2 \rightarrow \ep(z) \, \eta_2, &
\psi_{m2} \rightarrow \ep(z) \, \psi_{m2}, &
\psi_{51} \rightarrow  \ep(z) \, \psi_{51} \\[2mm]
{} & q_3  \rightarrow \ep(z) \, q_3. &{}\\
\ema
\eea
From now on, we work with the even parts of the odd fields.  We
assume that odd bosonic fields vanish on the brane,
\be
e_m^\hf=e_5^a=B_m=0 \quad {\rm on\ \Si.}
\label{bc}
\ee
This implies $e_5=e_4 e_5^\hf$ and $F_{m5}=\p_m B_5-\ep\p_5 B_m$ on $\Si$.  The even parts of the odd fermionic fields do not necessarily vanish on the brane.

We take the following ansatz for the brane action
\bea
\label{bracan}
S_B&=&\int d^5x e_4\da(z)\Big\{
-3\la_1-2\al_1\psi_{m1}\si^{mn}\psi_{n1}
-c_1\frac{i}{2}\chibar\sibar^m\Dh_m\chi\nn\\
&&\qquad -c_2\chi\chi
-6i c_3\psi_{m1}\si^m\chibar
-c_4e_5^\hf F^{m5}\chi\si_m\chibar
+\rm{h.c.}
\Big\} .
\eea
Equation (\ref{bracan}) includes a brane tension, $T_1=-6\la_1$, necessary to generate a warped background, together with a mass-like term for the gravitino $\psi_{m1}$ (see ref.\ \cite{bb1}).  The action also includes kinetic and interaction terms for a brane-localized fermion $\chi$.   The coefficients $\al_1$ and $c_i$ are arbitrary complex numbers.  We make $c_3$ real by a phase rotation on $\chi$.

Because of the brane action, the equations of motion for $e_m^a$ and $\psi_{m1}$ have terms proportional to $\da(z)$.  They cancel provided 
\be
\label{bcom}
\om_{ma\hf}=\ep\la_1 e_{ma}
\ee
and
\be
\label{bcpsim}
\psi_{m2}=\al_1\psi_{m1}-i c_3\si_m\chibar
\ee
on $\Si$, respectively.  The brane action also induces singular terms in the equations of motion for the bosonic fields appearing in $e_5^\hf F^{m5}$.  These terms are proportional to $\chi\si_m\chibar$ and can be neglected in the approximation we use.\footnote{We use the ``linearized supersymmetry'' approximation, in which we neglect two-Fermi terms in most bosonic quantities, three-Fermi terms in the supersymmetry transformations, and four-Fermi terms in the action.}

The above boundary conditions must be preserved under supersymmetry.  The variations of $e_m^\hf=0$ and $e_5^a=0$ in (\ref{bc}) imply
\be
\label{loclor1} 
\da e_m^\hf+e_m^a\om_a{}^{\hf}=0, \quad
\da e_5^a+e_5^\hf\om_\hf{}^{a}=0 \quad {\rm on\ \Sigma},
\ee
where we include a compensating local Lorentz rotation with parameter $\om^{a\hf}=-\om^{\hf a}$.  This requires
\bea
\label{bceta}
\eta_2 &=& \al_1\eta_1 \\
\label{bcpsi5}
\psi_{51}&=&-\al_1^\ast\psi_{52}-c_3 e_5^\hf \chi \\
\om^{a\hf}&=&-i c_3\eta_1\si^a\chibar+\rm{h.c.}\quad {\rm on\ \Sigma}.
\eea
(The compensating local Lorentz rotation was not necessary in ref.~\cite{bb1}; accordingly, $\om^{a\hf}$ vanishes when $\chi=0$.)

In the approximation we are using, the Goldstone fermion shifts under supersymmetry,
\be
\label{dachi}
\da\chi=v_1\eta_1 .
\ee
Equation (\ref{bcpsim}) is preserved under supersymmetry if
\be
\la_1(1+\aone\aone^\ast)+\la\left(
\aone\qstar+\aone^\ast q_{12}+(\aone\aone^\ast -1)q_3\right) 
= c_3 v_1^\ast,
\ee
where $q_{12}=q_1+i q_2$ and we have set $\ep^2=1$. This equation also implies that $v_1$ is real (since $c_3$ is real).

The condition $B_m=0$ in (\ref{bc}) is a little more subtle.  The variation
\be
\da B_m=\ha{\sqrt{6}}c_3\eta_1\si_m\chibar+{\rm{h.c.}}
=\da\left(\ha{\sqrt{6}}\frac{c_3}{v_1}\chi\si_m\chibar\right).
\ee
implies that, on the brane, $B_m$ is given in terms of a bilinear in $\chi$.  In our approximation, this is consistent with $B_m=0$.  Finally, invariance of (\ref{bcom}) and (\ref{bcpsi5}) under supersymmetry gives boundary conditions for $\p_5 \psi_{m2}$ and $\p_5 \eta_2$, respectively.  They will not be important in our discussion.

With these results, we are ready to compute the supersymmetry variation of the bulk-plus-brane action.  There are three contributions.  The first comes from the bulk action, resulting from the redefinition $q_3\rightarrow\ep(z)q_3$.  It is\footnote{The word ``same'' stands for the terms present in ref.~\cite{bb1}, where $\chi=0$.}
\bea
\da^{(1)}S_5 &=&\int d^5x e_4\da(z)\left[
6\la q_3(\ep^2\psi_{m2}\si^m\etabar_2-\psi_{m1}\si^m\etabar_1)+\rm{h.c.}
\right]\nn\\
&=&\int d^5x e_4\da(z)\left[
{\rm{same}}+ 24\la q_3\ep^2 c_3\al_1\chi\eta_1 +\rm{h.c.}\right].
\eea
The second comes from the modification of the transformation for $\psi_{52}$,
\be
\da\psi_{52}=\da\psi_{52}\Big|_{\rm{old}}-4\eta_2\da(z),
\ee
necessary to close the supersymmetry algebra and make $\da\psi_{52}$ finite on the brane.  It is
\bea
\da^{(2)}S_5&=&\int d^5x e_4\da(z)\big[
8\psi_{m1}\si^{mn}\wh D_n\eta_2
+6i\la(\ep^2 q_3\psi_{m2}+q_1\psi_{m1})\si^m\etabar_2
\nn\\
&& \qquad\qquad
-2i\om_{ma\hf}\eta_2(\si^{mn}\si^a+\si^a\si^{mn})\psi_{n2}
-\sx i e_5^\hf F^{m5}\eta_2\psi_{m1}+\rm{h.c.}
\big]\nn\\
&=&\int d^5x e_4\da(z)\left[
{\rm{same}}+24(\la_1+\la q_3)\ep^2 c_3\al_1\chi\eta_1+\rm{h.c.}\right].
\eea
The third contribution comes from the variation of the brane action, using (\ref{dachi}) together with the induced supersymmetry transformations,
\be
\da e_m^a = {\rm{same}} +\ep^2 c_3\al_1\eta_1\si^a\sibar_m\chi
+{\rm{h.c.}}, \quad \da\psi_{m1} =\rm{same}.
\ee
It is
\bea
\da S_B&=&\int d^5x e_4\da(z)\big\{{\rm{same}}
+(12c_3-v_1 c_1)i\chibar\sibar^m\wh D_m\eta_1
\nn\\&& \qquad
+(\sx c_3-2v_1 c_4)e_5^\hf F^{m5}\eta_1\si_m\chibar
-6i v_1 c_3\psi_{m1}\si^m\etabar_1
\nn\\[5pt]&&
+\left[24c_3(\la(q_{12}+\ep^2\al_1 q_3)+2\la_1\ep^2\al_1)
-2v_1 c_2\right]\chi\eta_1
+\rm{h.c.}\big\}
\eea
Adding all three contributions together, we find the variation of the full bulk-plus-brane action,
\bea
&&\da^{(1)}S_5+\da^{(2)}S_5+\da S_B =\int d^5x e_4\da(z)\Big[
C_1\psi_{m1}\si^{mn}\wh D_n\eta_1+i C_2\chibar\sibar^m\wh D_m\eta_1\nn \\
&&\qquad+\ e_5^\hf F^{m\hf}(i C_3\psi_{m1}\eta_1+C_4\eta_1\si_m\chibar)
+C_5\chi\eta_1 +6i C_6\psi_{m1}\si^m\etabar_1+\rm{h.c}\Big],
\eea
where
\bea
&&C_1=0, \quad C_2=-v_1 c_1+12c_3, \quad
C_3=0, \quad C_4=\sqrt{6}c_3-2v_1 c_4 \nn\\
&&C_5=-2v_1c_2+24c_3(3\al_1\ep^2(\la_1+\la q_3)+\la q_{12})\nn\\
&&C_6=-c_3v_1+\la_1(1+3\aone\aone^\ast \ep^2)+
\la(\aone\qstar+\aone^\ast q_{12}+(3\aone\aone^\ast \ep^2-1)q_3) . 
\eea
Therefore, the total action is supersymmetric if parameters in the brane action (\ref{bracan}) satisfy
\bea
&&c_3=\frac{1}{12}c_1v_1, \quad
c_2=c_1(\la q_{12}+\al_1(\la_1+\la q_3)), \quad
c_4=\frac{c_1}{4\sqrt{6}}\\
&&\la_1(1+\aone\aone^\ast)+
\la(\aone\qstar+\aone^\ast q_{12}+(\aone\aone^\ast-1)q_3)
=\frac{1}{12}c_1v_1^2 .
\eea
We have used $\ep^2(z)\da(z)=\frac{1}{3}\da(z)$, as usual.  The only undetermined parameter is $c_1 \in \mathbb{R}$. It is fixed by the normalization of the kinetic term for $\chi$ in the brane action.

\subsection{Two branes}

The above derivation can be readily extended to the case of a two-brane system in which there are two independent Goldstone fermions, $\chi_1(x)$ and $\chi_2(x)$, living on $\Si_1$ and $\Si_2$, respectively.  Using arguments like those above, it is not hard to show that the following two-brane action,
\bea
\label{twobrac}
S_B&=&+\int_{\Si_1}d^4x\Big\{
-3\la_1-2\al_1\psi_{m1}\si^{mn}\psi_{n1}
+\xi_1\Big[
-\ha{i}\chibar_1\sibar^m\Dh_m\chi_1\nn\\
&&\quad
-\ \ha{m_1}\chi_1\chi_1-\ha{i}v_1\psi_{m1}\si^m\chibar_1
-\frac{1}{4\sqrt{6}}e_5^\hf F^{m5}\chi_1\si_m\chibar_1
\Big]+{\rm{h.c.}}
\Big\}\nn \\
&&
-\int_{\Si_2}d^4x\Big\{
-3\la_2-2\al_2\psi_{m1}\si^{mn}\psi_{n1}
+\xi_2\Big[
-\ha{i}\chibar_2\sibar^m\Dh_m\chi_2 \nn\\
&&\quad
-\ \ha{m_2}\chi_2\chi_2-\ha{i}v_2\psi_{m1}\si^m\chibar_2
-\frac{1}{4\sqrt{6}}e_5^\hf F^{m5}\chi_2\si_m\chibar_2
\Big]+\rm{h.c.}
\Big\},
\eea
with parameters
\bea
m_{1,2}&=&2\left[\la q_{12}+\al_{1,2}(\la_{1,2}+\la q_3)\right]\nn\\[2mm]
\label{vlaal}
\xi_{1,2} v_{1,2}^2&=&12\left[
\la_{1,2}(1+\al_{1,2}\al_{1,2}^\ast)+\la(\al_{1,2}^\ast q_{12}
+\al_{1,2}\qstar+q_3(\al_{1,2}\al_{1,2}^\ast-1))\right],
\eea
is consistent with local supersymmetry in the full bulk-plus-brane system.  The relative minus sign between actions for $\Si_1$ and $\Si_2$ is convenient because
\be
\ep^\pr(z)=2\da_1(z)-2\da_2(z) ,
\ee 
where $\da_{1,2}(z)$ are delta functions corresponding to the locations of the branes, $\Si_{1,2}$.  The kinetic terms for $\chi_1$ and $\chi_2$ are properly normalized and have the correct signs if $\xi_1=1$ and $\xi_2=-1$; opposite signs correspond to ghost-like fields.

Closure of the supersymmetry algebra and invariance of the bulk-plus-brane action is achieved if
\begin{enumerate}
\item
the supersymmetry transformations for the induced fields in the brane action are exactly the ones induced from the bulk;
\item
the Goldstone fermions living on the branes transform as follows,
\be
\da\chi_{1,2}=v_{1,2}\eta_1;
\ee
\item
the supersymmetry variations satisfy the boundary conditions on $\Si_{1,2}$,
\be
\label{bceta12}
\eta_2=\al_{1,2}\eta_1;
\ee 
\item
the bulk fields satisfy the following boundary conditions on $\Si_{1,2}$,
\bea
&&e_m^\hf=e_5^a=B_m=0\\ 
\label{bcom12}
&&\om_{ma\hf}=\ep\la_{1,2} e_{ma}\\
\label{bcpsim12}
&&\psi_{m2}=\al_{1,2}\psi_{m1}-\frac{i}{12}\xi_{1,2} v_{1,2}\si_m\chibar_{1,2}\\
\label{bcpsi512}
&&\psi_{51}=-\al_{1,2}\psi_{52}-\frac{1}{12}\xi_{1,2}v_{1,2}e_5^\hf\chi_{1,2};
\eea
\item
the supersymmetry transformation for $\psi_{52}$ is modified by
\be
\da\psi_{52}=\da\psi_{52}\Big|_{\rm{old}}
-4(\al_1\da_1(z)-\al_2\da_2(z))\eta_1,
\ee
thus making it non-singular on the branes;
\item
the supersymmetry variations are accompanied by field-dependent local Lorentz rotations, $\da e_M^A=e_M^B\om_B{}^A$, with
\be
\om^{ab}=0, \quad 
\om^{a\hf}=-\frac{i}{12}\xi_{1,2}v_{1,2}\eta_1\si^a\chibar_{1,2}+\rm{h.c.}\quad{\rm on\ \Si_{1,2}.}
\ee
\end{enumerate}
Without the Goldstone fermions, these conditions are just those of ref.~\cite{bb1}.  

Note that the presence of the Goldstone fermions relaxes the condition $|\la_{1,2}|\leq\la$. Indeed, when $\xi_{1,2}\neq 0$, we can choose $\la_{1,2}$ and $\al_{1,2}$ arbitrarily; eq.~(\ref{vlaal}) then gives the required $v_{1,2}$.  Accordingly, in the effective four-dimensional theory, the cosmological constant can now be positive, as expected when supersymmetry is spontaneously broken on the branes.


\section{Dimensional reduction}
\label{DimRed}

In our construction, the Goldstone fermions $\chi_{1,2}$ are localized on the branes.  The spontaneous supersymmetry breaking is communicated to the bulk fields through the boundary conditions.  In this section we carry out a consistent dimensional reduction of the bulk-plus-brane action down to four dimensions.  In this way we show how the bulk-plus-brane action determines the supersymmetry breaking in the effective theory.

For simplicity, we make the following assumptions:
\begin{enumerate}
\item 
All parameters and warp factors are real.\footnote{Note that for Scherk-Schwarz supersymmetry breaking one must consider {\it complex} $\al_{1,2}$; see ref.~\cite{bb2}. Here we do not mix the two ways of supersymmetry breaking.}  In particular, for the unit vector $\vec q$ we assume $q_2=0$, so that $q_{12}=q_1$ and $q_1^2+q_3^2=1$.
\item
The radion multiplet is frozen.
\end{enumerate}
We start by presenting the effective action, then we carry out the bosonic and fermionic reductions.

\subsection{Effective action}

In general, the effective theory has three light spin-half fields:  a Goldstone fermion plus two others.  The Goldstone fermion is a linear combination of $\chi_1$, $\chi_2$, and the superpartner of the radion.  In our reduction, we freeze the radion and ignore both non-Goldstone linear combinations of the spin-half fields.  The resulting four-dimensional low-energy supergravity theory describes the four-dimensional veirbein $\wh e_m^a(x)$, the gravitino $\psi_m(x)$, together with a Goldstone fermion $\G(x)$.  It is given by the action
\bea
\label{4action}
S =\int d^4x \wh e_4 \Big\{
-\ha{1}\wh R +3g^2-\ha{1} v^2
+\big[
\ha{1}\wh \eps^{mpnk}\psibar_m\wh\sibar_p\wh D_n\psi_k
-\ha{i}\Gbar\wh\sibar_m\wh D^m \G \nn\\
-(g\psi_m\wh\si^{mn}\psi_n+\ha{i}v\psi_m\wh\si^m\Gbar
+g\G\G)+\rm{h.c.}\big]\Big\} .
\eea
The action is invariant under the following (nonlinear) supersymmetry transformations,
\bea
\label{4ste}
&&\da\wh e_m^a =i\eta\si^a\psibar_m +\text{h.c.} \nn\\
\label{4stm}
&&\da\psi_m=2\wh D_m\eta+i g\wh\si_m\etabar \nn\\
\label{4stg}
&&\da \G=v\eta .
\eea
In what follows we find a reduction that consistently takes the five-dimensional bulk-plus-brane action and supersymmetry transformations into the ones given above.

\subsection{Ansatz}

We start with the following ansatz relating the five-dimensional fields to their four-dimensional counterparts,
\bea
&&e_m^a=a(y)\wh e_m^a, \quad e_5^\hf=1, \quad
e_\hf^a=e_m^\hf=B_m=B_5=0 \nn\\
&&\eta_1=\beta_1(y)\eta, \quad
\psi_{m1}=\beta_1(y)\psi_m+i\nu_1(y)\wh\si_m\Gbar, \quad
\psi_{51}=\rho_1(y)\G \nn\\
&&\eta_2=\beta_2(y)\eta, \quad
\psi_{m2}=\beta_2(y)\psi_m+i\nu_2(y)\wh\si_m\Gbar, \quad
\psi_{52}=\rho_2(y)\G .
\eea
Here all the $x^5=z$-dependence is separated into the warp factors; the dimensionless coordinate $y=\la|z|$. We define dimensionless parameters $g_0$ and $v_0$ by writing $g=\la g_0$ and $v=\la v_0$.

\subsection{Bosonic reduction}

The reduction of the bosonic part of the bulk-plus-brane action is the same as the one given in ref.~\cite{bb1}. The only difference is that we now write the effective cosmological constant as\footnote{Recall that $\Lambda_4$ arises as a separation constant when passing from the five-dimensional Einstein equations to the four-dimensional ones.} 
\be
\Lambda_4=-3g^2+\frac{v^2}{2}.
\ee
Therefore the bosonic warp factor must satisfy the following bulk equations
\be
\label{eqay}
a^{\pr\pr}(y)=a, \quad 
a^\pr(y)^2=a^2-g_0^2+\frac{v_0^2}{6}
\ee
and boundary conditions, 
\be
\label{bcay}
\la_i=-\la\frac{a^\pr(y)}{a(y)}\Big|_{\Si_i},
\ee
following from eq.~(\ref{bcom12}).  With these restrictions, the bosonic part of the bulk-plus-brane action,
\be
S_{5B}=\int d^5x e_5\Big[
-\frac{1}{2}R+6\la^2\Big]
-6 \la_1 \int_{\Si_1}d^4x e_4
+6 \la_2 \int_{\Si_2}d^4x e_4
\ee
reduces to
\be
S_{5B} =\oint dz a^2\int d^4x \wh e_4 \Big[
-\frac{1}{2}\wh R -\Lambda_4\,\Big],
\ee
as required.  The overall integral over $a^2$ renormalizes the gravitational coupling constant in four dimensions.

\subsection{Fermionic reduction. Part 1}

In the previous section, we used the equations of motion to determine the bosonic warp factor.  In this section, we use the supersymmetry transformations to find the fermionic warp factors.  This procedure is not precisely equivalent to the reduction of the equations of motion; we discuss the difference in appendix \ref{KK}.  The two approaches, however, lead to the same effective action.

Let us first consider the supersymmetry transformations of the bosonic fields.  We require 
\be
\da e_m^a+e_{mb}\om^{ba}=a\da\wh e_m^a, \quad
\da e_m^\hf+e_{ma}\om^{a\hf}=0, \quad
\da e_5^a+\om^{\hf a}=0, \quad \da e_5^\hf=0,
\ee
where we have included a compensating Lorentz rotation.  We find
\bea
&&\om^{a\hf}=i(\beta_1\rho_1+\beta_2\rho_2)\eta\si^a\Gbar+\text{h.c.}\\
&&a\om^{ab}=-2(\beta_1\nu_1+\beta_2\nu_2)\eta\si^{ab}\G+\text{h.c.}\\
\label{b+b}
&&\beta_1^2+\beta_2^2=a\\
\label{b-n}
&&\beta_1\nu_2-\beta_2\nu_1=a(\beta_1\rho_1+\beta_2\rho_2),
\eea
where the first equation, the next two, and the last one follow from the conditions on $\da e_5^a$, $\da e_m^a$, and $\da e_m^\hf$, respectively.  The remaining non-vanishing terms in $\da e_m^a$, $\da e_5^\hf$, $\da B_m$ and $\da B_5$ are variations of fermionic bilinears, using eq.~(\ref{4stg}).  Consistency with the supersymmetry transformations is achieved if we change the ansatz as follows,
\bea
e_m^a &=& a\wh e_m^a+\frac{1}{2v}(\beta_1\nu_1+\beta_2\nu_2)
(\G\G+\text{h.c.})\wh e_m^a \nn\\
\frac{2}{\sqrt6}B_m &=& 0+\frac{1}{v}(\beta_2\nu_1-\beta_1\nu_2)
\G\wh\si_m\Gbar\nn\\
e_5^\hf+i\frac{2}{\sqrt6}B_5 &=& 1+\frac{1}{v}\G\G .
\eea
Since we are working in an approximation where we neglect such fermionic bilinears, we leave this modification implicit.

Let us now consider the fermionic fields.  We require
\bea
&&\da\psi_{51}=\rho_1\da \G, \quad
\da\psi_{m1}=\beta_1\da\psi_m+i\nu_1\wh\si_m\da\Gbar\nn\\
&&\da\psi_{52}=\rho_2\da \G, \quad
\da\psi_{m2}=\beta_2\da\psi_m+i\nu_2\wh\si_m\da\Gbar.
\eea
This leads to the following equations,
\bea
\label{rb1}
&&v_0\rho_1=2\beta_1^\pr-(q_1\beta_2-q_3\beta_1)\\
\label{rb2}
&&v_0\rho_2=2\beta_2^\pr-(q_1\beta_1+q_3\beta_2)\\
\label{nb1}
&&v_0\nu_1=a(q_1\beta_1+q_3\beta_2)-a^\pr\beta_2-g_0\beta_1\\
\label{nb2} 
&&v_0\nu_2=-a(q_1\beta_2-q_3\beta_1)+a^\pr\beta_1-g_0\beta_2 .
\eea
One can easily check that eq.~(\ref{b-n}) follows from these four and eq.~(\ref{b+b}). Therefore, we have only five equations for six fermionic warp factors, $\beta_{1,2}$, $\nu_{1,2}$ and $\rho_{1,2}$.   We will see that one more equation follows from the reduction of the fermionic part of the bulk-plus-brane action.

\subsection{Fermionic reduction. Part 2}

In this section we find one additional equation for the warp factors, coming from the diagonalization of the kinetic terms for the fermions in the effective action.  This will complete the system of equations that determines all the warp factors.

Using the ansatz expressions for $e_5^\hf$, $e_\hf^a$, $e_m^5$, $B_m$ and $B_5$, we can write the fermionic part of the bulk-plus-brane action as follows (see ref.~\cite{bb1}),
\be
\label{total5}
S_{5F}=S_{5F}^{\not{\,\da}}+S_{5F}^\da,
\ee
where
\bea
\label{fermi-da}
S_{5F}^{\not{\,\da}}&=&\int d^5x e_4 \Big\{\ha{1}\eps^{mpnk}
(\psibar_{m1}\sibar_p D_n\psi_{k1}+\psibar_{m2}\sibar_p D_n\psi_{k2})\nn\\
&&
+\ \psi_{m1}\si^{mn}\ep\p_5\psi_{n2}-\psi_{m2}\si^{mn}\ep\p_5\psi_{n1}
+2(\psi_{52}\si^{mn}D_m\psi_{n1}-\psi_{51}\si^{mn}D_m\psi_{n2})\nn\\
&&
-\ \ha{3}\la q_1(\psi_{m1}\si^{mn}\psi_{n1}-\psi_{m2}\si^{mn}\psi_{n2}
    +i\psi_{m1}\si^m\psibar_{52}+i\psi_{m2}\si^m\psibar_{51})\nn\\
&&
-\ \ha{3}\la q_3(2\psi_{m1}\si^{mn}\psi_{n2}+i\psi_{m2}\si^m\psibar_{52}
    -i\psi_{m1}\si^m\psibar_{51})
+\text{h.c.}\Big\}
\eea
and
\bea
\label{fermi+da}
S_{5F}^\da&=&\int d^5x e_4\Big\{
2(\da_1(z)-\da_2(z))\psi_{m1}\si^{mn}\psi_{n2}
-2(\al_1\da_1(z)-\al_2\da_2(z))\psi_{m1}\si^{mn}\psi_{n1}\nn\\
&&
+\ \xi_1\da_1(z)\big[
-\ha{i}\chibar_1\sibar^m\wh D_m\chi_1-\ha{m_1}\chi_1\chi_1
-\ha{i}v_1\psi_{m1}\si^m\chibar_1\big]\nn\\
&&
-\ \xi_2\da_2(z)\big[
-\ha{i}\chibar_2\sibar^m\wh D_m\chi_2-\ha{m_2}\chi_2\chi_2
-\ha{i}v_2\psi_{m1}\si^m\chibar_2\big] 
+\text{h.c.}\Big\}.
\eea
All the brane-localized fermionic terms are in $S_{5F}^\da$.  Note that the first term in $S_{5F}^\da$ comes from the bulk action after the redefinition $\psi_{m2} \rightarrow \ep\psi_{m2}$.

Using $e_m^a=a\wh e_m^a$, together with its consequences,
\be
\si_m=a\wh\si_m, \quad
\si^{mn}=a^{-2}\wh\si^{mn}, \quad
\eps^{mpnk}=a^{-4}\wh\eps^{mpnk}, \quad
e_4=a^4\wh e_4 ,
\ee
and the ansatz expressions for $\psi_{m1,2}$ and $\psi_{51,2}$, one can reduce $S_{5F}^{\not{\,\da}}$ to the following form,
\bea
\label{5acA}
S_{5F}^{\not{\,\da}}&=&\int d^4x\wh e_4\oint dz a^2\Big\{
\ha{1} A_1\wh\eps^{mpnk}\psibar_m\wh\sibar_p\wh D_n\psi_k
+3A_2 i\Gbar\wh\sibar{}^m\wh D_m\G
+2A_3\G\wh\si^{mn}\wh D_m\psi_n\nn\\
&&\qquad \qquad \qquad 
+\la A_4\psi_m\wh\si^{mn}\psi_n
-\ha{3}\la A_5 i\psi_m\wh\si^m\Gbar
+6\la A_6\G\G +\text{h.c.}\Big\},
\eea
where 
\bea
&&A_1=\frac{1}{a}(\beta_1^2+\beta_2^2) \nn\\
&&A_2=\frac{1}{a}(\nu_1^2+\nu_2^2)-(\rho_2\nu_1-\rho_1\nu_2)\nn\\
&&A_3=\beta_1\rho_2-\beta_2\rho_1-\frac{2}{a}(\beta_1\nu_1+\beta_2\nu_2)\nn\\
&&A_4=\beta_1\beta_2^\pr-\beta_2\beta_1^\pr-\ha{3}\big[
(\beta_1^2-\beta_2^2)q_1+2\beta_1\beta_2 q_3\big]\nn\\
&&A_5=\beta_1\nu_2^\pr-\beta_2\nu_1^\pr+\nu_1\beta_2^\pr-\nu_2\beta_1^\pr
+\big[(\beta_1\rho_2+\beta_2\rho_1)q_1+(\beta_2\rho_2-\beta_1\rho_1)q_3\big]a\nn\\
&&\qquad \quad -(\beta_1\rho_1+\beta_2\rho_2)a^\pr
-3\big[(\beta_1\nu_1-\beta_2\nu_2)q_1+(\beta_1\nu_2+\beta_2\nu_1)q_3\big]\nn\\
&&A_6=\nu_1\nu_2^\pr-\nu_2\nu_1^\pr
+\big[(\nu_1\rho_2+\nu_2\rho_1)q_1+(\nu_2\rho_2-\nu_1\rho_1)q_3\big]a\nn\\
&&\qquad \quad -(\nu_1\rho_1+\nu_2\rho_2)a^\pr
-\ha{3}\big[(\nu_1^2-\nu_2^2)q_1+2\nu_1\nu_2 q_3\big].
\eea
The term with $A_3$ must be eliminated to have the usual kinetic terms for $\psi_m$ and $\G$.  This gives the sixth equation for the fermionic warp factors,
\be
\label{sixtheq}
\beta_1\nu_1+\beta_2\nu_2=\frac{a}{2}(\beta_1\rho_2-\beta_2\rho_1).
\ee

We now have what we need to find all the warp factors. First, we must solve eqs.~(\ref{eqay}) and (\ref{bcay}) to find $a(y)$.  We then find $\beta_1(y)$ and $\beta_2(y)$ from the following system of equations
\be
\label{fweqs}
\beta_1^2+\beta_2^2=a, \quad
\beta_1\beta_2^\pr-\beta_2\beta_1^\pr=\ha{3}\big[
(\beta_1^2-\beta_2^2)q_1+2\beta_1\beta_2 q_3\big]-g_0
\ee
with boundary conditions
\be
\label{bcbeta}
\al_i=\frac{\beta_2(y)}{\beta_1(y)}\Big|_{\Si_i} ,
\ee
following from eq.~(\ref{bceta12}). The remaining warp factors, $\rho_{1,2}(y)$ and $\nu_{1,2}(y)$, can then be simply calculated using eqs.~(\ref{rb1})--(\ref{nb2}).

\subsection{Fermionic reduction. Part 3}
In this section we complete the reduction of the fermionic action.  We will see that the singular part of the bulk-plus-brane action does not vanish; it plays a crucial role in the reduction.

In preparation for what follows, it is convenient to introduce
\be
\tau_1 \equiv \beta_1\rho_1+\beta_2\rho_2, \quad
\tau_4 \equiv \beta_1\rho_2-\beta_2\rho_1.
\ee
Using the system of equations for warp factors, we find
\bea
\label{tau1g2}
v_0\tau_1 
&=& a^\pr-[2\beta_1\beta_2 q_1-(\beta_1^2-\beta_2^2)q_3]\\
\label{tau4g1}
v_0 \frac{\tau_4}{2}
&=& -g_0+[(\beta_1^2-\beta_2^2)q_1+2\beta_1\beta_2 q_3].
\eea
The warp factors $\rho_{1,2}(y)$ and $\nu_{1,2}(y)$ can be expressed in terms of $\tau_1$ and $\tau_4$ as 
\bea
&&a\rho_1=\beta_1\tau_1-\beta_2\tau_4, \quad 
\nu_1=\beta_1\ha{\tau_4}-\beta_2\tau_1 \\
&&a\rho_2=\beta_1\tau_4+\beta_2\tau_1, \quad
\nu_2=\beta_1\tau_1+\beta_2\ha{\tau_4} .
\eea
It takes only some algebra to check that the coefficients $A_i$ can be simplified to 
\bea
&&A_1=1, \quad A_3=0, \quad A_4=-g_0 \nn\\
&&A_2=T, \quad A_5=\frac{v_0}{6}-v_0 T, \quad A_6=g_0 T,
\eea
where
\be
T=2\tau_1^2-\frac{\tau_4^2}{4}.
\ee
If $T$ were equal to $-1/6$, the action $S_{5F}^{\not{\,\da}}$ would be identical with the fermionic part of the action (\ref{4action}).  However, this is not the case.  The matching requires a contribution from the singular part of the action,\footnote{In ref.~\cite{bb1}, the singular part of the bulk-plus-brane action vanished once boundary conditions were applied. This does not happen in general.} eq.~(\ref{fermi+da}).

Applying the boundary condition for $\psi_{m2}$, eq.~(\ref{bcpsim12}), we can write the singular part of the fermionic action, eq.~(\ref{fermi+da}), as follows,
\bea
\label{f+da}
&&S_{5F}^\da=\int d^5x e_4\Big\{
\xi_1\da_1(z)\big[
-\ha{i}\chibar_1\sibar^m\wh D_m\chi_1-\ha{m_1}\chi_1\chi_1
-\frac{i}{4}v_1\psi_{m1}\si^m\chibar_1\big]\nn\\
&&\qquad
-\xi_2\da_2(z)\big[
-\ha{i}\chibar_2\sibar^m\wh D_m\chi_2-\ha{m_2}\chi_2\chi_2
-\frac{i}{4}v_2\psi_{m1}\si^m\chibar_2\big] 
+\text{h.c.}\Big\}.
\eea
With the help of the ansatz, the boundary conditions (\ref{bcpsim12}), (\ref{bcpsi512}) and (\ref{bcbeta})  imply
\bea
\label{psi5chi1}
&&\tau_1(z_1)\G(x)=-\frac{\xi_1 v_1}{12}\beta_1(z_1)\chi_1(x)\\
\label{psi5chi2}
&&\tau_1(z_2)\G(x)=-\frac{\xi_2 v_2}{12}\beta_1(z_2)\chi_2(x),
\eea
where $z_{1,2}$ correspond to the locations of the branes $\Si_{1,2}$.
Applying a supersymmetry variation to these equations, we find\footnote{Equations (\ref{tau1xi1}) and (\ref{tau1xi2}) can also be derived from the bulk equation (\ref{tau1g2}), the boundary conditions (\ref{bcay}) and (\ref{bcbeta}), and the relation between parameters, eq.~(\ref{vlaal}). This is one of many consistency checks in this dimensional reduction.}
\bea
\label{tau1xi1}
&&\tau_1(z_1)v = -\frac{\xi_1 v_1^2}{12}\beta_1^2(z_1)\\
\label{tau1xi2}
&&\tau_1(z_2)v = -\frac{\xi_2 v_2^2}{12}\beta_1^2(z_2).
\eea
Using these relations, we can cast eq.~(\ref{f+da}) into the following form
\bea
S_{5F}^\da &=& \int d^4x \wh e_4 \oint dz a^3 (\da_1(z)-\da_2(z))\Big\{
6\tau_1\frac{1}{v} i\Gbar\wh\sibar{}^m\wh D_m\G 
\nn\\
&&\qquad
+3\tau_1 i\psi_m\wh\si^m\Gbar
+12\tau_1\frac{g}{v}\G\G
+\text{h.c.}\Big\} .
\eea

Let us now rewrite this singular contribution as a contribution to the bulk action, using
\be
\oint dz(\da_1(z)-\da_2(z))f(z)=-\ha{1}\oint dz\ep(z)\wt f{}^\pr(z)=
-\ha{\la}\oint dz \wt f{}^\pr(y),
\ee
where $\wt f(z)$ is any function that equals $f(z)$ at $z_{1,2}$.  In this way we absorb $S_{5F}^\da$ into $S_{5F}^{\not{\,\da}}$, correcting the values of the $A_i$,
\bea
&&a^2\wt A_2=a^2 A_2-\frac{1}{v_0}(a^3\tau_1)^\pr\nn\\
&&a^2\wt A_5=a^2 A_5+(a^3\tau_1)^\pr\nn\\
&&a^2\wt A_6=a^2 A_6-\frac{g_0}{v_0}(a^3\tau_1)^\pr .
\eea
It is now only the matter of algebra to prove that
\be
\wt A_2=-\frac{1}{6}, \quad
\wt A_5=\frac{v_0}{3}, \quad
\wt A_6=-\frac{g_0}{6}.
\ee
These are precisely the values necessary to match the effective four-dimensional action (\ref{4action}).

We see that our ansatz, together with the equations and boundary conditions for the warp factors, consistently reduces the original five-dimensional bulk-plus-brane system to four-dimensional supergravity, described by the veirbein $\wh e_m^a(x)$, the gravitino $\psi_m(x)$ and a Goldstone fermion $\G(x)$.

One comment, however, is in order.  Equations (\ref{psi5chi1}) and (\ref{psi5chi2}) imply that $\G(x)$ is proportional to $\chi_1(x)$ and $\chi_2(x)$.  Our reduction requires that $\chi_1$ and $\chi_2$ be identified, up to multiplicative constants.  This follows from the fact that we have assumed there is only one spin-half fermion in the effective action.  A more general reduction would be able to accommodate independent $\chi_1$ and $\chi_2$, together with the superpartner of the radion field.

Note that our reduction is also sufficient for the case when a Goldstone fermion is present on only one of the two branes.  Equations (\ref{tau1xi1}) and (\ref{tau1xi2}) guarantee that if $\xi_i v_i=0$, then $\tau_1(z_i)=0$, and the corresponding relation between $\G$ and $\chi_i$ is eliminated.


\section{Example: Randall-Sundrum scenario}
\label{Example}

In this section we will illustrate the reduction for the Randall-Sundrum scenario, where $\Si_1$ (the Planck brane) and $\Si_2$ (the TeV brane) are at $z_1=0$ and $z_2=\pi R$, respectively, with tensions tuned to satisfy $\la_1=\la_2=\la$.  In this case, the effective theory has zero cosmological constant, $\Lambda_4=0$.  We choose $q_3=1$ (so $q_1=q_2=0$).  The brane action is therefore eq.~(\ref{twobrac}) with\footnote{Note that in the tuned case, eq.~(\ref{ghost}) requires $\xi_2\geq0$. This means that $\chi_2$ (the Goldstone fermion on the negative tension brane) is a ghost-like field on $\Si_2$.} 
\bea
&&\la_1=\la, \quad m_1=4\la\al_1, \quad \xi_1 v_1^2=24\la\al_1^2 \\
\label{ghost}
&&\la_2=\la, \quad m_2=4\la\al_2, \quad \xi_2 v_2^2=24\la\al_2^2 .
\eea

The bosonic warp factor, normalized to unity on the Planck brane, $a(0)=1$, is
\be
a(y)=\exp(-y) .
\ee
Since $\Lambda_4=0$, we must have
\be
\label{gisv}
|g|=\frac{|v|}{\sx}.
\ee

The fermionic warp factors, $\beta_1$ and $\beta_2$, are given by
\be
\beta_1^2=\frac{a}{1+u^2}, \quad \beta_2=u\beta_1,
\ee
where $u(y)$ is a solution to 
\be
u^\pr(y)=3u-g_0\frac{1+u^2}{a} ,
\label{ueq}
\ee
subject to the following boundary conditions,
\be
u(0)=\al_1, \quad u(\la\pi R)=\al_2.
\ee
The solution is
\be
\label{usolution}
u(y)=\frac{J_1(g_0\exp(y))+s Y_1(g_0\exp(y))}{
J_2(g_0\exp(y))+s Y_2(g_0\exp(y))},
\ee
where
\be
-s=\frac{J_1(g_0)-\al_1 J_2(g_0)}{Y_1(g_0)-\al_1 Y_2(g_0)}
=\frac{J_1(g_0\exp(\la\pi R))-\al_2 J_2(g_0\exp(\la\pi R))}
{Y_1(g_0\exp(\la\pi R))-\al_2 Y_2(g_0\exp(\la\pi R))}
\label{bcrs}
\ee
because of the boundary conditions. The last relation, together with eq.~(\ref{gisv}), implicitly determines the scale of supersymmetry breaking, $v$, in terms of the brane parameters and the proper distance between the branes.

We now obtain a solution for $g=g(\al_1,\al_2,\la\pi R)$ under certain simplifying assumptions.  The Kaluza-Klein masses for the gravitino scale as $m_n \approx n\pi\la\exp(-\la\pi R)$ in the absence of supersymmetry breaking \cite{pom}.  If we assume that the supersymmetry breaking mass shift, $g$, is much smaller than the first Kaluza-Klein mass, and that $\la\pi R$ is sufficiently large, as necessary to generate a hierarchy, we find
\be
g_0 \ll \exp(-\la\pi R) \ll 1.
\ee
In this approximation, eq.~(\ref{bcrs}) reduces to
\be
-s\approx\frac{\pi g_0^3}{4(2\al_1-g_0)}
\approx\frac{\pi g_0^3 e^{3\la\pi R}}{4(2\al_2-g_0 e^{\la\pi R})}
\ee
and therefore
\be
g\approx 2\la(\al_1-\al_2 e^{-3\la\pi R}).
\ee

Let us now restore the gravitational coupling constants $k_5$ and $k_4$ so that
\be
k_5\da\chi_1=v_1\eta_1, \quad
k_5\da\chi_2=v_2\eta_1, \quad
k_4\da \G=v\eta.
\ee
By a field redefinition, we can always choose $v_{1,2}$ and $v$ to be positive.  Using
\be
\frac{k_5^2}{k_4^2}=\oint dz a^2=2\int_0^{\pi R}dz e^{-2\la z}
=\frac{1}{\la}(1-e^{-2\la\pi R})\approx \frac{1}{\la},
\ee
and substituting for $\alpha_1$ and $\alpha_2$, we find
\be
g\approx\frac{k_4}{k_5\sx}
(\sqrt{\xi_1}v_1-\sqrt{\xi_2}v_2 e^{-3\la\pi R}).
\ee
This is equivalent to
\be
\frac{v}{k_4}\approx
\left|
\sqrt{\xi_1}\frac{v_1}{k_5}-\sqrt{\xi_2}\frac{v_2}{k_5}e^{-3\la\pi R}
\right| .
\ee

We conclude by focussing to two possible choices:
\begin{enumerate}
\item
Goldstone fermion on Planck brane, $(\xi_1,\xi_2)=(1,0)$.
\be
\da\chi_1=\wh v_1{}^2\eta_1 \quad \Rightarrow \quad
\da \G=\wh v{\,}^2\eta, \quad
\wh v{\,}^2\approx \wh v_1{}^2 .
\ee
The scale of supersymmetry breaking is transmitted full strength to the effective  theory.
\item
Goldstone fermion on the TeV brane, $(\xi_1,\xi_2)=(0,1)$.
\be
\da\chi_2=\wh v_2{}^2\eta_1 \quad \Rightarrow \quad
\da \G=\wh v{\,}^2\eta, \quad
\wh v{\,}^2\approx \wh v_2{}^2 e^{-3\la\pi R} .
\ee
The effective scale of supersymmetry breaking is exponentially suppressed compared to the contribution from the hidden brane.  Note that in this case, the Goldstone fermion $\chi_2$ is a ghost-like field on $\Sigma_2$. However, the Goldstone fermion of the effective theory, $\G$, is not ghost-like.
\end{enumerate}


\section{Conclusion}

In this paper we study brane-localized supersymmetry breaking in the five-dimensional Randall-Sundrum scenario.  Our analysis is model independent; our only assumption is that nonlinearly transforming Goldstone fermions live on the branes.  Our results can be applied to any model in which supersymmetry is spontaneously broken by physics localized on one (or more) of the branes.

We pay great attention to the boundary conditions, which come from matching singular terms in equations of motion and requiring consistency with (local) supersymmetry.  The boundary conditions ensure that supersymmetry is broken {\it spontaneously}, rather than explicity by a mismatch between bosonic and fermionic boundary conditions.  Furthermore, our analysis was {\it not} done in a fixed bosonic background (as is the case in most of the work on this subject; see, e.g., ref.~\cite{mp}).  As the result, our results are directly applicable to any background, subject to the boundary conditions.

Our work extends the results of ref.~\cite{bb1}, where it was shown that the Randall-Sundrum scenario can be supersymmetrized not only in the tuned case \cite{first}, $\la_1=\la_2=\pm\la$, but also in the detuned case, provided the tensions satisfy the bound $|\la_{1,2}|<\la$.  With Goldstone fermions on the branes, the bulk-plus-brane system can be made locally supersymmetric even when $|\la_{1,2}|>\la$.  This gives rise to a dS$_4$ bosonic background in which global supersymmetry is necessarily broken.

Our analysis holds for general $\vec q=(q_1, 0, q_3)$, which shows that there is no essential difference between the two ``orthogonal'' choices \cite{bkvp}, even when coupling
to matter is present. The choice $q_3=1$, however, is often more convenient for calculations.
We leave the full Kaluza-Klein reduction of our construction for future research, in particular, the derivation of the effective theory for independent Goldstone fermions, $\chi_1(x)$ and 
$\chi_2(x)$, interacting with the radion multiplet.

\vspace{0.2in}

This work was supported in part by the U.S. National Science Foundation, grant NSF-PHY-9970781.


\appendix
\section{Notation}
We define the following dimensionless quantities,
\be
y=\la|z|, \quad
g_0=\frac{g}{\la}, \quad
v_0=\frac{v}{\la} .
\ee
We set $k_5=k_4=1$; they can be restored by rescaling
\bea
&&S_5\rightarrow k_5^2 S_5, \quad
B_{m,5}\rightarrow k_5 B_{m,5}, \quad
\psi_{m1,2}\rightarrow k_5\psi_{m1,2}, \quad
\chi_{1,2}\rightarrow k_5\chi_{1,2},\nn\\
&&S_4\rightarrow k_4^2 S_4, \quad
\psi_{m}\rightarrow k_4\psi_{m}, \quad
\G \rightarrow k_4\G .
\eea
The mass dimensions of certain parameters and fields are collected in Table 1.

\begin{table}[t]
\begin{center}
\begin{tabular}{|c|c|c|c|c|c|c|c|}\hline
$-3/2$ & $-1$ & $-1/2$ & 0 & 1/2 & 1 & 3/2 & 2 \cr\hline
$k_5$ & $k_4$ & $\eta_{1,2}$ & $e_M^A$ & $v_{1,2}$ & $\la$,\ $\la_{1,2}$ 
& $\psi_m$ & $\psi_{m1,2}$\cr
{} & $z $& $\eta$ & $\wh e_m^a$ &  & $g$, $v$ & $\G$ & $\psi_{51,2}$  \cr
 & & & $\al_{1,2}$ & & $\wh v_{1,2}$ &$ \chi_{1,2}$ & \cr
 & & & $y$, $g_0$, $v_0$ & & $\da(z)$ & $B_m,\ B_5$ &\cr\hline
\end{tabular}
\caption{Mass dimensions of parameters and fields.}
\end{center}
 \end{table}
 

\section{Warp factors}

To find explicit solutions for the warp factors, one can proceed as follows. Given the constants $g$ and $v$, one finds $a(y)$, satisfying
\be
a^{\pr\pr}(y)=a, \quad 
a^\pr(y)^2=a^2-g_0^2+\frac{v_0^2}{6}.
\ee
The solution is simple for both positive and negative cosmological constant, $\Lambda_4=-3g^2+\dst \textstyle \frac{1}{2} v^2 $.  One then solves
\be
\beta_1^2+\beta_2^2=a, \quad
\beta_1\beta_2^\pr-\beta_2\beta_1^\pr=\ha{3}\big[
(\beta_1^2-\beta_2^2)q_1+2\beta_1\beta_2 q_3\big]-g_0,
\ee
to find $\beta_1(y)$ and $\beta_2(y)$, and uses equations (\ref{rb1})--(\ref{nb2}) to compute the other warp factors.  The boundary conditions 
\be
\la_{1,2}=-\la\frac{a^\pr}{a}\Big|_{\Si_{1,2}}, \quad 
\al_{1,2}=\frac{\beta_2}{\beta_1}\Big|_{\Si_{1,2}}
\ee
fix the integration constants and restrict the input parameters.

The calculation can be simplified by introducing  $u=\beta_2/\beta_1$. Then 
\be
\beta_1^2=\frac{a}{1+u^2}, \quad \beta_2=u\beta_1,
\ee
and for a given $a(y)$, one solves a single equation,
\be
\label{equ13}
u^\pr=\ha{3}\big[(1-u^2)q_1+2u q_3\big]-g_0\frac{1+u^2}{a},
\ee
with boundary condition 
\be
u\big|_{\Si_{1,2}}=\al_{1,2}.
\ee

It is sufficient to solve eq.~(\ref{equ13}) for a specific choice of parameters $(q_1, q_3)$.  For example, if $u_3(y)$ is a solution for $(q_1, q_3)=(0,1)$, then 
\be
u=\frac{1-q_3+q_1 u_3}{(1-q_3)u_3-q_1}
\ee
is a solution for any other combination $(q_1, q_3)$. The equation for $u_3(y)$ is just
\be
u_3^\pr=3u_3-g_0\frac{1+u_3^2}{a}.
\ee

The solution can be written explicitly in three special cases:
\begin{enumerate}
\item
$v=0$. Pure anti-de Sitter, $\Lambda_4=-3g^2$.
\be
a(y)=g_0\cosh(y), \quad u_3(y)=e^{y}
+\frac{2c e^{3y}}{(1+e^{2y})^2-c(1+2e^{2y})}.
\ee 
\item
$g=0$. Pure de Sitter, $\Lambda_4=\dst\ha{v^2}$.
\be
a(y)=\frac{v_0}{\sqrt6}\sinh(y), \quad u_3(y)=c e^{3y}.
\ee
\item
$g=\dst\frac{v}{\sx}$. Minkowski, $\Lambda_4=0$.
\be
a(y)=\exp(\mp y), \quad 
u_3(y)=\left[
\frac{J_1(g_0 e^{\pm y})+c Y_1(g_0 e^{\pm y})}
{J_2(g_0 e^{\pm y})+c Y_2(g_0 e^{\pm y})}
\right]^{\pm 1} .
\ee
\end{enumerate}
Here $J_n$ and $Y_n$ are Bessel functions, and $c$ is a free parameter.  Another free parameter arises from shifting $y$ by a constant.


\section{Difference from the standard KK reduction}
\label{KK}

The standard Kaluza-Klein reduction for the warped two-brane scenario was carried out in ref.~\cite{pom}.  The equations for the fermionic warp factors, eqs.~(15) and (16) of ref.~\cite{pom}, do not agree with ours, eq.~(\ref{fweqs}) of this paper.  Nevertheless, the {\it ratio} of the warp factors is the same (our eq.~(\ref{usolution}) and eqs.~(17) and (18) in ref.~\cite{pom}).  In this appendix we explain the discrepancy.  Before we can do that, however, we need to sketch how the dimensional reduction proceeds using the fermionic equations of motion.

The five-dimensional fermionic equations of motion follow from the bulk action (\ref{fermi-da}),
\bea
\frac{\da S}{\da\psibar_{m1,2}} &=&
-\ \eps^{mpnk}\si_p D_n\psibar_{k1,2}
\pm 2\si^{mn}(D_5\psi_{n2,1}-D_n\psi_{52,1}) 
\nn\\
&&\mp\ 3\la\si^{mn}(q_1\psi_{n1,2}\pm q_3\psi_{n2,1})
-\frac{3}{2}\la i\si^m(q_1\psibar_{52,1}\mp q_3\psibar_{52,1})
\ =\ 0 .
\eea
Using our ansatz
\bea
e_m^a=a(y)\wh e_m^a, \quad
\psi_{m1,2}=\beta_{1,2}(y)\psi_m+i\nu_{1,2}(y)\wh\si_m\Gbar, \quad
\psi_{51,2}=\rho_{1,2}(y)\G ,
\eea
and remembering that
\be
D_n\psi_{m1,2}=\wh D_m\psi_{m1,2} 
\pm \frac{i}{2}\la\om\si_m\psibar_{m2,1}, \quad
\om \equiv -\frac{a^\pr}{a},
\ee
we rewrite the equations of motion as follows,
\bea
a^3\frac{\da S}{\da\psibar_{m1,2}} &=&
-\beta_{1,2}\wh\eps{}^{mpnk}\wh\si_p\wh D_n\psibar_k
+2(2\nu_{1,2}\mp a\rho_{2,1})\wh\si^{mn}\wh D_n\G 
\nn\\
&& \pm 2\la a K_{1,2} \wh\si^{mn}\psi_n
\mp 3\la a N_{1,2} i\wh\si^m\Gbar=0,
\eea
where
\bea
K_{1,2} &=&
\beta_{2,1}^\pr-
(\om\beta_{2,1}+\frac{3}{2}(q_1\beta_{1,2}\pm q_3\beta_{2,1}))
\nn\\
N_{1,2} &=&
\nu_{2,1}^\pr
-(\om\nu_{2,1}+\frac{3}{2}(q_1\nu_{1,2}\pm q_3\nu_{2,1})
\pm\frac{1}{2}a(\om\rho_{1,2}+q_1\rho_{2,1}\mp q_3\rho_{1,2})) .
\eea

To proceed further, we require that the four-dimensional fields satisfy the following four-dimensional equations of motion,
\bea
&&\wh\eps{}^{mpnk}\wh\si_p\wh D_n\psibar_k =
-2g\wh\si^{mn}\psi_n
-\frac{i}{2}v\wh\si^m\Gbar
\nn\\
&&\wh D_m\G =
\frac{v}{2}\psi_m
-\frac{i}{2}g\wh\si_m\Gbar+\phi_m,
\eea
where
\bea
\wh\si^m\phibar_m=0.
\eea
Then the five-dimensional equations of motion reduce to
\bea
a^3\frac{\da S}{\da\psibar_{m1,2}} &=&
2(2\nu_{1,2}\mp a\rho_{2,1})\wh\si^{mn}\phi_n
\pm 2\la a \wt K_{1,2} \wh\si^{mn}\psi_n
\mp 3\la a \wt N_{1,2} i\wh\si^m\Gbar\ =\ 0,
\eea
where
\bea
&& \wt K_{1,2} = K_{1,2} 
\pm\frac{g_0}{a}\beta_{1,2}
\pm\frac{v_0}{2a}(2\nu_{1,2}\mp a\rho_{2,1})
\nn\\
&& \wt N_{1,2} = N_{1,2}
\mp\frac{v_0}{6a}\beta_{1,2}
\mp\frac{g_0}{2a}(2\nu_{1,2}\mp a\rho_{2,1}) .
\eea

Next, we use the following equations for the warp factors,
\bea
&&a^{\pr\pr}= a, \quad 
(a^\pr)^2= a^2-g_0^2+\frac{v_0^2}{6}
\nn\\[2mm]
&&v_0\rho_{1,2}=2\beta_{1,2}^\pr-(q_1\beta_{2,1}\mp q_3\beta_{1,2})
\nn\\
&&v_0\nu_{1,2}=\beta_{1,2}(\pm a q_1-g_0)+\beta_{2,1}(a q_3\mp a^\pr) ,
\eea
together with $q_1^2+q_3^2=1$. These relations imply
\bea
\wt K_{1,2}=0, \quad
\wt N_{1,2}=0.
\eea
Accordingly, the five-dimensional equations of motion are simply
\bea
a^3\frac{\da S}{\da\psibar_{m1,2}} &=&
2(2\nu_{1,2}\mp a\rho_{2,1})\wh\si^{mn}\phi_n\ =\ 0.
\eea
Note that these equations hold for {\it arbitrary} $\beta_{1,2}(y)$.\footnote{Note also that if we contract the five-dimensional equations of motion with $\wh\si_m$, we find zero because of the condition $\si^m\phibar_m=0$. Therefore, the contracted equations are satisfied for {\it arbitrary} $\beta_{1,2}(y)$. A similar calculation shows that equations of motion obtained from varying $\psi_{51,2}$ are also satisfied for arbitrary $\beta_{1,2}(y)$.}

If we require the five-dimensional equations to be satisfied point-by-point along the fifth dimension, we are led to the following relation,
\bea
\label{KKnus}
2\nu_{1,2}\mp a\rho_{2,1} =0.
\eea
This puts constraints on $\beta_{1,2}(y)$,
\bea
\label{KKbetas}
\beta_{1,2}^\pr
-\frac{3}{2}(q_1\beta_{2,1}\mp q_3\beta_{1,2})
-\om\beta_{1,2}
=\pm\frac{g_0}{a}\beta_{2,1} .
\eea
For $q_3=1$ and $a(y)=\exp(-y)$ (the tuned case, with $\la_1=\la_2=\la$), these equations match precisely (with obvious identifications) eqs.~(15) and (16) of ref.~\cite{pom}.  In this case the solution is
\bea
\beta_{1,2}(y) = N \exp(\frac{3}{2}y)
\left( J_{2,1}(g_0\exp(y))+ s Y_{2,1}(g_0\exp(y)) \right),
\eea
and therefore the ratio of $\beta_2$ and $\beta_1$ coincides with the expression in eq.~(\ref{usolution}).

In general, though, the $\beta_{1,2}$ obtained from eq.~(\ref{KKbetas}) differ from the $\beta_{1,2}$ of section 3.  This can be seen from the fact that the main relation between the bosonic and fermionic warp factors, eq.~(\ref{b+b}),
\be
\beta_1^2+\beta_2^2=a,
\ee
cannot be satisfied simultaneously with eqs.~(\ref{KKbetas}) when ${a^\pr}^2=a^2-g_0^2+v_0^2/6$ unless $v_0=0$ (in which case the $N=1$ supersymmetry is linearly realized and we are back to the case considered in ref.~\cite{bb1}).

In this paper we sacrifice the five-dimensional fermionic equations of motion in favor of point-by-point reduction of the supersymmetry transformations.  We demand only that the five-dimensional action correctly reduce to its four-dimensional counterpart.  This requires that the following linear combination of the five-dimensional equations of motion,
\bea
\frac{\da S}{\da\psibar_m} =
\beta_1\frac{\da S}{\da\psibar_{m1}} +
\beta_2\frac{\da S}{\da\psibar_{m2}} ,
\eea
must vanish. Therefore, instead of two equations (\ref{KKnus}), there is only one,
\bea
\beta_1(2\nu_{1}-a\rho_{2})
+ \beta_2(2\nu_{2}+a\rho_{1}) =0 ,
\eea
which is equivalent to the following equation for $\beta_{1,2}$,
\be
\beta_1\beta_2^\pr-\beta_2\beta_1^\pr=\ha{3}\big[
(\beta_1^2-\beta_2^2)q_1+2\beta_1\beta_2 q_3\big]
-g_0\frac{\beta_1^2+\beta_2^2}{a}.
\ee
This equation is common for both approaches; it is the reason why the ratio of $\beta_2$ and $\beta_1$ turns out to be the same.

In our approach, the reduction of supersymmetry transformations works point-by-point, but the equations of motion are reduced only ``on average.''  Similarly, in the standard KK approach, the equations of motion are reduced point-by-point, but the supersymmetry transformations for the effective action require averaging over the fifth dimension.  The two approaches differ, but they lead to the same effective action.



\end{document}